# NbSeTe — A New Layered Transition Metal Dichalcogenide Superconductor

**Dong Yan[1], Shu Wang[1], Yishi Lin[2], Guohua Wang[3], Yijie Zeng[4], Mebrouka Boubeche[1], Yuan He[1], Jie Ma[3], Yihua Wang[2], Dao-Xin Yao[4], Huixia Luo[1*]**

[1]School of Material Science and Engineering and Key Lab of Polymer Composite & Functional Materials, Sun Yat-Sen University, No. 135, Xingang Xi Road, Guangzhou, 510275, P. R. China

[2]Department of Physics, Fudan University, Shanghai, 200433, China

[3]Department of Physics and Astronomy, Shanghai Jiao Tong University, Shanghai 2 002 40, China

[4]School of Physics, Sun Yat-Sen University, No. 135, Xingang Xi Road, Guangzhou, 510275, P. R. China

* Corresponding author: H.X. Luo. E-mail address: luohx7@mail.sysu.edu.cn

## ABSTRACT

Transition metal dichalcogenides (TMDC's) usually exhibit layered polytypic structures due to the weak interlayer coupling. 2H-NbSe$_2$ is one of the most widely studied in the pristine TMDC family due to its high superconducting transition temperature ($T_c$ = 7.3K) and the occurrence of a charge-density wave (CDW) order below 33 K. The coexistence of CDW with superconductivity poses an intriguing open question about the relationship between Fermi surface nesting and Cooper pairing. Past studies of this issue have mostly been focused on doping 2H-NbSe$_2$ by 3$d$ transition metals without significantly changing its crystal structure. Here we replaced the Se by Te in 2H-NbSe$_2$ in order to design a new 1T polytype layered TMDC NbSeTe, which adopts a trigonal structure with space group $P\bar{3}m1$. We successfully grew large size and high-quality single crystals of 1T-NbSeTe *via* the vapor transport method using I$_2$ as the transport agent. Temperature-dependent resistivity and specific heat data revealed a bulk $T_c$ at 1.3 K, which is the first observation of superconductivity in pure 1T-NbSeTe phase. This compound enlarged the family of superconducting TMDC's and provides an opportunity to study the interplay between CDW and superconductivity in the trigonal structure.

# INTRODUCTION

Transition metal dichalcogenides (TMDCs) have been widely investigated due to their diverse structural and electronic properties such as large thermoelectric effect, anomalous magnetoresistance, charge-density-wave (CDW), spin-charge-orbital density waves (SDW), and superconductivity[1-10]. The layered TMDCs' structures are generally made from the stacking two-dimensional X-T-X sandwiches (T is typically a transition metal atom from group IVB, VB, VIB and X is S, Se, or Te atoms). Each sandwich layer is bonded through covalent and weak van der Waals (vdWs) force. The layered TMDCs usually have *6R, 4H$_a$, 4H$_b$, 3R, 2H, 1T* phases[1,2], where the integer indicates the number of X-T-X sandwiches per unit cell perpendicular to the layers, while R, H, and T denote rhombohedral, hexagonal, trigonal symmetries, respectively. These layered TMDC structures can easily be broken and rebuilt via substitution or intercalation due to the week connects force between layers.

TMDCs with 2H (trigonal prismatically coordinated) and 1T (octahedrally coordinated) structures have been extensively studied for several years due to the correlation between their charge density wave (CDW) state and superconducting property. 2H-NbSe$_2$ is a famous TMDCs material, which possess a superconducting state with $T_c \sim 7.3$ K and quasi-two-dimensional incommensurate CDW state with $T_{ICDW} \sim 33$ K simultaneously.[11] Besides, the $T_c$ of 2H-NbSe$_2$ is remarkably higher than those of other superconducting TMDC materials, where $T_c$ is usually in the 2 - 4 K range. Thus, 2H-NbSe$_2$ becomes the most abundantly researched of the layered TMDC superconductors. Up until now, numerous theoretical and experimental works about its physical properties have been reported. The researchers prefer to use the effective and mature methods to regulate and control the CDW and the superconducting behavior, such as chemical doping [12-18], adopting high pressure [19-22], gating [23] and controlling layers of the nanoscale samples. It has been established previously that the intercalation of 3$d$ transition metals (such as Zn, Al, Co, Ni, Fe) into 2H-NbSe$_2$ quickly destroyed the $T_c$ to < 1 K. [24-25] Recently, Cava group has also reported that the intercalation of Cu into 2H-NbSe$_2$ to obtain Cu$_x$NbSe$_2$ ($0 \leq x \leq 0.09$) [15]. However, they found out the type that $T_c$ is suppressed with doping in this situation is unusual, an *S*-shaped downtrend was found, with an inflection point near $x = 0.03$ and a leveling off of the $T_c$ near 3 K at higher doping content. The similar behavior is observed in our recently reported double doping $Cu_xNbSe_{2-y}S_y$ ($0 \leq x = y \leq 0.09$) system. [16] It has also been found that its superconductivity is suppressed by the substitution on Se site by Te atoms into 2H-NbSe$_2$ to form 2H-NbTe$_x$Se$_{2-x}$ ($x = 0, 0.10, 0.20$), where 2H polytype was maintained. [26] These previously reported works all maintained 2H-NbSe$_2$ structure no matter intercalation or substitution.

As we all know that the physical properties and electronic structures of TMDCs are highly dependent on the crystal structures. One significant property of layered TMDCs is its highly polymorphic and polytypism character. [27-28] Previously, we have found that TaSe$_{2-x}$Te$_x$ for ($0 \leq x \leq 2$) can be crystallized in 1T, 2H, 3R and monoclinic structure (1T') polytypes, where the 2H, 3R and 1T polymorphs are all superconducting while the monoclinic phase shows normal metallic behavior to 0.4 K. By far, although many efforts have been focusing on 2H-polytype NbSe$_2$, no sufficient research attention has been paid to the 1T form.

In this article, we first report the synthesis, structure determination, and superconducting characterization of 1T-NbSeTe. Single crystals of 1T-NbSeTe have been successfully synthesized *via* chemical vapor transport method. The structure was confirmed by single crystal and powder X-ray diffraction measurements, which reveal that the obtained 1T-NbSeTe sample crystallized in the trigonal structure with space group $P\text{-}\bar{3}m1$. The resistivity and heat capacity measurements were performed to characterize 1T-NbSeTe samples systematically, which reveal that 1T-NbSeTe is a type-II superconductor and has a $T_c \approx 1.3$ K. An obvious sharp peak at $T_c$ in the heat capacity shows that the normalized specific-heat jump $\Delta C/\gamma T_c \approx 1.36$, which was very approximate to the predicted value by weak-coupling BCS theory (1.43). This compound enlarged the family of superconducting TMDCs and opens an opportunity to study the interplay between polytypism and superconducting property of the transition metal dichalcogenides.

# EXPERIMENTAL SECTION

The polycrystalline specimens of 1T-NbSeTe were obtained *via* a solid-state method. Firstly, we weigh the reactants of Te (99.9%), Se (99.999%), and Nb (99.999%) in stoichiometric ratios and ground well in the agate mortar. Then the mixture was transferred in sealed vacuum silica glass tubes. The polycrystalline samples were got after heating the tubes at 850 ºC for 120 hours. In order to ensure sample uniformity, we reground, re-pelletized, and sintered as-prepared polycrystalline samples at 850 ºC for 48 hours. The single crystals of 1T-NbSeTe were grown *via* the chemical vapor transport (CVT) method, using $I_2$ as a transport agent. The as-prepared 1T-NbSeTe powders were mixed with $I_2$ at quality ratio 20:1, then heated in sealed vacuum silica tubes for 7 days in a two-zone furnace, where the source and growth zones temperatures were 800 °C and 700 °C, respectively.

Single crystal diffraction data of 1T-NbSeTe was collected by using SuperNova X-ray Source, Mo Kα/Cu radiation at 293 K. The data sets and absorption were corrected with the Lorentz and polarization factors and Multiscan method respectively.[28] The structure was solved by direct methods and refined by a full-matrix least-squares fitting on $F^2$ by SHELX-97.[30] We used the PLATON program to check possible missing symmetry for the structure, but none was found.[31] Crystal data and the structural refinements information of 1T-NbSeTe are reported in Table 1. Important Bond Lengths (Å) for 1T-NbSeTe are presented in Table S2.

We also adopted the powder X-ray diffraction (PXRD, Bruker D8 focus, Cu Kα radiation, graphite diffracted beam monochromator) to characterize the samples structurally. The unit cell parameters were determined by profile fitting the powder diffraction data with FULLPROF diffraction suite with Thompson-Cox-Hastings pseudo-Voigt peak shapes.[32] A Quantum Design Physical Property Measurement System (PPMS) equipped with a $^3$He was used to take the measurements of the temperature dependence of electrical resistivity and heat capacity from 0.4 K to 300 K. No materials' air-sensitivity was found during the measurements. $T_c$ was determined from the midpoint of the resistivity $\rho$(T) transitions, for the specific heat data, we employed equal area construction method to obtain the critical temperature.

The composition and microscopic morphology of 1T-NbSeTe samples were investigated by energy dispersive X-ray spectroscopy (EDXS) and scanning electron microscopy (SEM, Quanta 400F, Oxford), respectively. And high-resolution transmission electron microscopy (HRTEM, FEI Tecnai G2 F20 operated on 200 kV) was also used to study the phase structure of 1T-NbSeTe.

## RESULTS AND DISCUSSIONS

1T-NbSeTe crystallizes in the trigonal space group of $P\bar{3}m1$ with unit cell parameters of $a = b$ = 3.5707(12) Å, $c$ = 6.524(4)) Å, and Z = 1. In the asymmetric unit, two crystallographically independent positions are observed for Nb, mixed Se/Te atoms; one position for fully occupied Nb; another one for half occupied Se/Te. The crystal structure of 1T-NbSeTe is shown in **Fig. 1.** Slightly distorted, edge sharing, Nb(Se/Te)$_6$ octahedra form the chains that extend along the *b*-axis. Then these chains are linked by Se/Te atoms to form two-dimensional (2D) layers; the Nb-Se/Te distances are in the range of 2.7026(18) – 2.7027(18) Å and the *cis* Se/Te-Nb-Se/Te angles locates in the range of 82.69(6) ～ 97.31(7)º. All of the layers connect with each other *via* vdW forces and stack vertically along *c*-axis. Crystallographic data got *via* single crystal X-ray diffraction were presented in **Table 1**, final atomic parameters are summarized in **Table 2**. Powder X-ray diffraction data of 1T-NbSeTe are presented in **Fig. 2a**. All site of the peaks consists well with the structure data of single crystal obtained early. What's more, powder diffractometer was used to determine the *c* direction in crystal morphology through the flat face of single crystal on a glass slide, predominantly (00l) peaks are illustrated in **Fig. 2b**, which indicates that the *c* direction is perpendicular along the blade-shaped sample. The defective alignment on glass slide of the crystal probably leads to minor peaks in other directions an imperfect alignment.

**Fig. 3c** is the scan electron microscope (SEM) images of single crystals 1T-NbSeTe,

which confirm that a layered high quality single crystal was growth successfully. The high-resolution transmission electron microscope (HRTEM) image (**Fig. 3b**) of NbSeTe powder samples shows clear (001) lattice distance of 0.652 nm. Fast Fourier transform images (FFT) are shown in the upper right corner of the image. **Fig. 3a** shows the [001] zone axis pattern (ZAP) of 1T-NbSeTe. It indicates that 1T-NbSeTe has a trigonal structure; which is consistent with the XRD data previously mentioned. The X-ray energy dispersive spectroscopy (EDS) spectrum (**Fig. 3d**) shows three elements of Nb, Se and Te in the 1T-NbSeTe single crystals with element ratio of Nb, Se and Te close to the formula 1:1:1, further indicating that high-quality single crystal with the composition of 1T-NbSeTe is obtained.

The temperature dependence electrical resistivity ($\rho(T)$) in the temperature range from 0.4 – 300K under zero magnetic field for polycrystalline 1T-NbSeTe sample, was presented in the main part of the **Fig. 4**. The sample reveals a metallic temperature dependence (d$\rho$/d$T$) in the temperature region of 2 to 300 K. The inset of **Fig. 4** presents the d$\rho$/d$T$ vs $T$ curve under low temperature (blue line), which further confirms $T_c$ is around 1.3 K. A sharp drop of $\rho(T)$ obviously takes place below 2.0 K (inset of **Fig. 4** (red line)), which represents the occurrence of superconductivity and demonstrates the details of the superconducting transition.

For the sake of getting more abundant information of the electronic and superconducting properties of 1T-NbSeTe polycrystalline sample, we took the heat capacity measurements for the polycrystalline sample under 0 T and 1 T applied magnetic field. The obtained $T_c$ from heat capacity measurements matched the $T_c$ obtained *via* temperature dependence electrical resistivity, which provides convincing evidence that a bulk superconducting state is achieved in 1T-NbSeTe. By fitting the normal state data of specific heats at high temperature under 1 T magnetic fields, it is found to obey the formula $C_p = \gamma T + \beta T^3$, where $\gamma$ and $\beta$ describe the electronic and phonon contributions to the heat capacity, respectively. We obtained the electronic specific-heat coefficients $\gamma = 5.7$ mJ mol$^{-1}$ K$^{-2}$ and phonon specific-heat coefficients $\beta = 0.8$ mJ mol$^{-1}$ K$^{-4}$ (inset of **Fig. 5**) from fitting the data gotten under 1 T applied field in temperature range of 2 – 4 K. Using the equal-area method [44], we estimate $\Delta C/T_c$ = 7.98 mJ mol$^{-1}$ K$^{-2}$ and the calculated normalized specific heat jump value $\Delta C/\gamma T_c$ (**Fig. 5**) was 1.36 for 1T-NbSeTe, which was very closed to the Bardeen-Cooper-Schrieffer (BCS) weak-coupling limit value (1.43), confirming bulk superconductivity. Then we obtain the Debye temperature *via* the formula $\Theta_D = (12\pi^4 nR/5\beta)^{1/3}$ by using the fitted value of $\beta$, where $n$ is the number of atoms per formula unit and $R$ is the gas constant. Therefore, the electron-phonon coupling constant ($\lambda_{ep}$) is calculated using the Debye temperature ($\Theta_D$) and critical temperature $T_c$ from the inverted McMillan formula: $\lambda_{ep} = \frac{1.04 + \mu^* \ln\left(\frac{\Theta_D}{1.45T_c}\right)}{(1-1.62\mu^*)\ln\left(\frac{\Theta_D}{1.45T_c}\right) - 1.04}$ [45]. This resultant $\lambda_{ep}$ is 0.55, suggesting that NbSeTe belongs to an intermediately coupled superconductor. The electron density of states at the Fermi level ($N(E_F)$) was obtained from the formula $N(E_F) = \frac{3}{\pi^2 k_B^2 (1+\lambda_{ep})}\gamma$ with the $\lambda_{ep}$ and $\gamma$. Then we got the value $N(E_F)$ = 1.56 states/eV f.u. for 1T-NbSeTe (**Table S1**).

Finally, we compared the superconducting transition temperatures ($T_c$s) of the common 1T-polytype superconducting TMDCs in **Fig. 6**. All of the $T_c$ data were obtained from temperature dependent resistivity. We can easily notice that $T_c$ of 1T-NbSeTe locates in the middle of 1T-polytype superconducting TMDCs' $T_c$ region. The results of this research fully indicates that substitution into polytype TMDCs not only expand the family of superconducting TMDCs, but also provide the opportunity to study the interplay between polytypism and superconductivity in the TMDCs.

**CONCLUSIONS**

A novel stoichiometric compound, 1T-NbSeTe, was synthesized, and its crystal structure and superconducting property were characterized. 1T-NbSeTe crystallizes in a trigonal space group of $P\bar{3}m1$ with unit cell parameters of $a = b$ = 3.5707(12) Å, $c$ = 6.524(4)) Å. 1T-NbSeTe shows two-dimensional layered structure composed of distorted Nb(Se/Te)$_6$ octahedra chains which are connected *via* Se/Te atoms. Resistivity measurements suggest that 1T-NbSeTe is a superconductor with transition temperature $T_c \approx$ 1.3 K. The heat capacity measurements confirm this property as well.

This compound 1T-NbSeTe enlarged the family of superconducting TMDCs and offers an opportunity to investigate the interplay between polytypism and superconductivity in the transition metal dichalcogenides.

**Supporting Information:** X–ray crystallographic files in CIF format, superconducting parameters in 1T-NbSeTe and important bond lengths (Å) for 1T-NbSeTe.

**Corresponding Author**

*FAX: (+86)20-39386124;* E-mail: luohx7@mail.sysu.edu.cn

**Notes**

The authors declare no competing financial interest.

# Acknowledgments

H. X. Luo acknowledges the financial support by "Hundred Talents Program" of the Sun Yat-Sen University and Natural Science Foundation of China (21701197). J. Ma is supported by the National Natural Science Foundation of China (Grant No. 11774223).

**Table 1.** Summary of crystallographic data and structure refinement parameters for 1T-NbSeTe at 293 K.

| Chemical Formula | **NbSeTe** |
|---|---|
| Formula weight | 299.47 |
| Crystal size (mm$^3$) | 0.100 × 0.050 × 0.020 |
| Crystal system | Trigonal |
| Space group | *P*-3*m*1 |
| *a* (Å) | 3.5707(12) |
| *b* (Å) | 3.5707(12) |
| *c* (Å) | 6.524(4) |
| *V* (Å$^3$) | 72.04(6) |
| Z | 1 |
| $D_{cal}$(g cm$^{-3}$) | 6.903 |
| $\mu$(mm$^{-1}$) | 26.375 |
| $\theta$ range (°) | 6.25-32.30 |
| GOF on $F^2$ | 0.998 |
| $R_1{}^a$ [$I > 2\sigma(I)$] | 0.0414 |
| $wR_2{}^b$ [$I > 2\sigma(I)$] | 0.0829 |
| $R_1{}^a$ (all data) | 0.0561 |
| $wR_2{}^b$ (all data) | 0.0880 |
| Flack parameter *x* | - |

$^a R_1 = \Sigma ||F_o| - |F_c||/\Sigma |F_o|$, $^b wR_2 = \Sigma[(w(F_o{}^2 - F_c{}^2)^2)/\Sigma[w(F_o{}^2)^2)]^{1/2}$

**Table 2.** Atomic coordinates and equivalent isotropic displacement parameters of 1T-NbSeTe

| Atom | Site | *Occ.* | *x* | *y* | *z* | $U_{eq}$ |
|---|---|---|---|---|---|---|
| Nb1 | 1*b* | 1.0 | 0.00000 | 1.00000 | 0.50000 | 0.052(1) |
| Se1 | 2*d* | 0.5 | 0.33333 | 0.66667 | 0.2321(4) | 0.031(1) |
| Te1 | 2*d* | 0.5 | 0.33333 | 0.66667 | 0.2321(4) | 0.031(1) |

$U_{eq}$ is defined as one-third of the trace of the orthogonalized $U_{ij}$ tensor (Å$^2$). The numbers in parentheses are the standard error of the refined parameters.

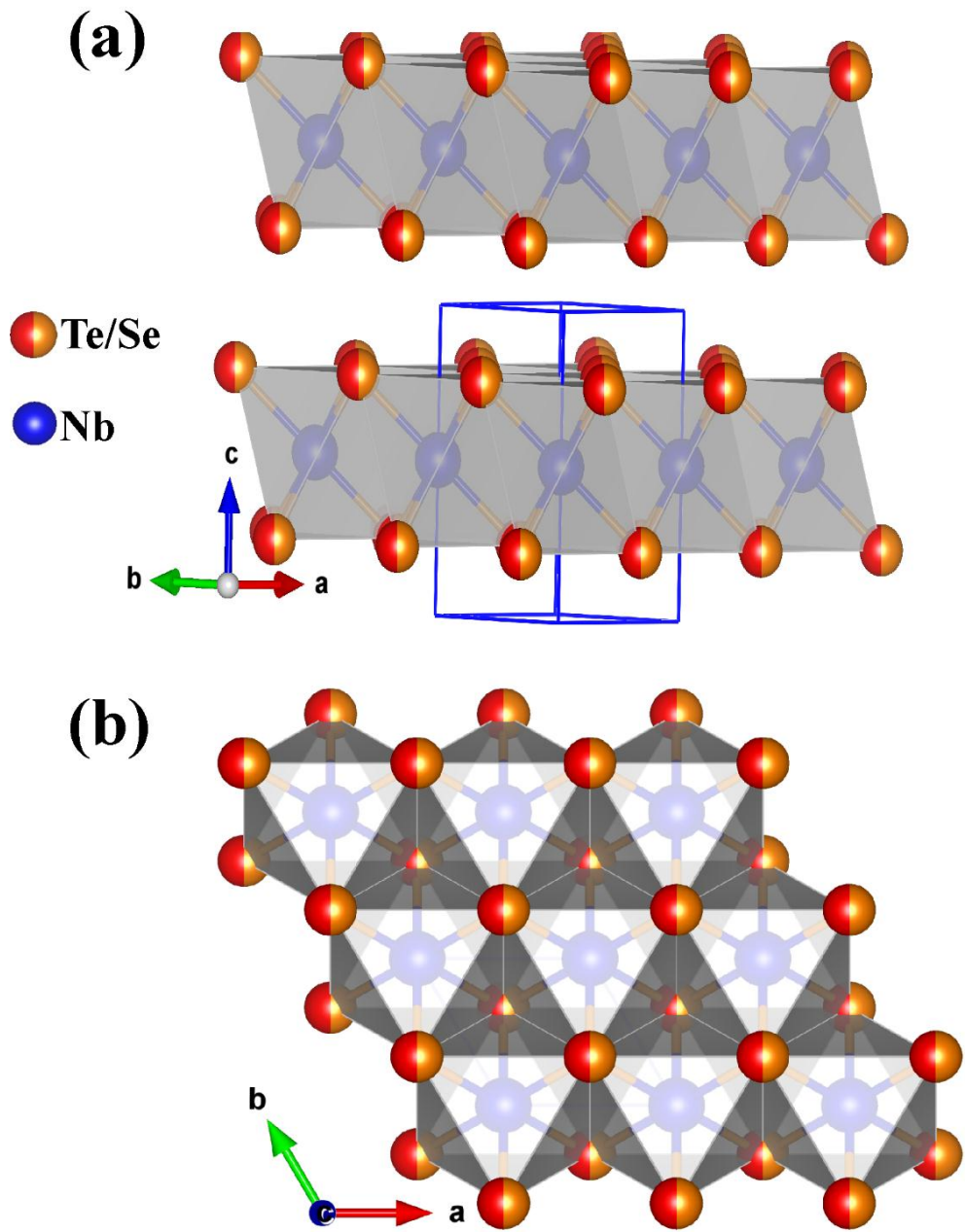

**Figure 1**. View of NbSeTe crystal structure along (111) direction (a), (001) direction (b).

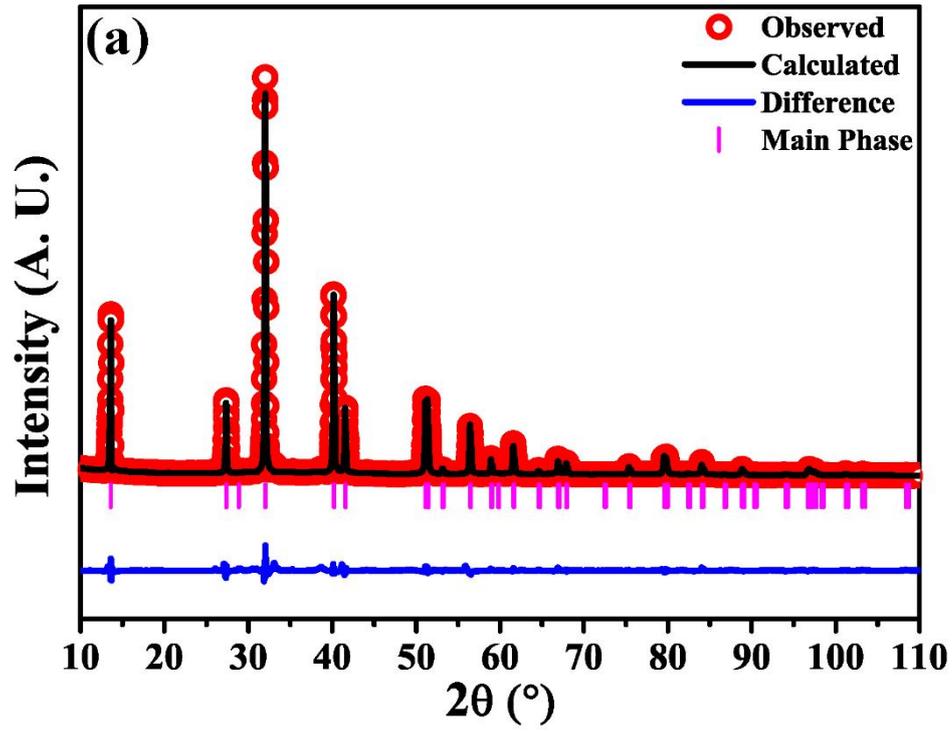

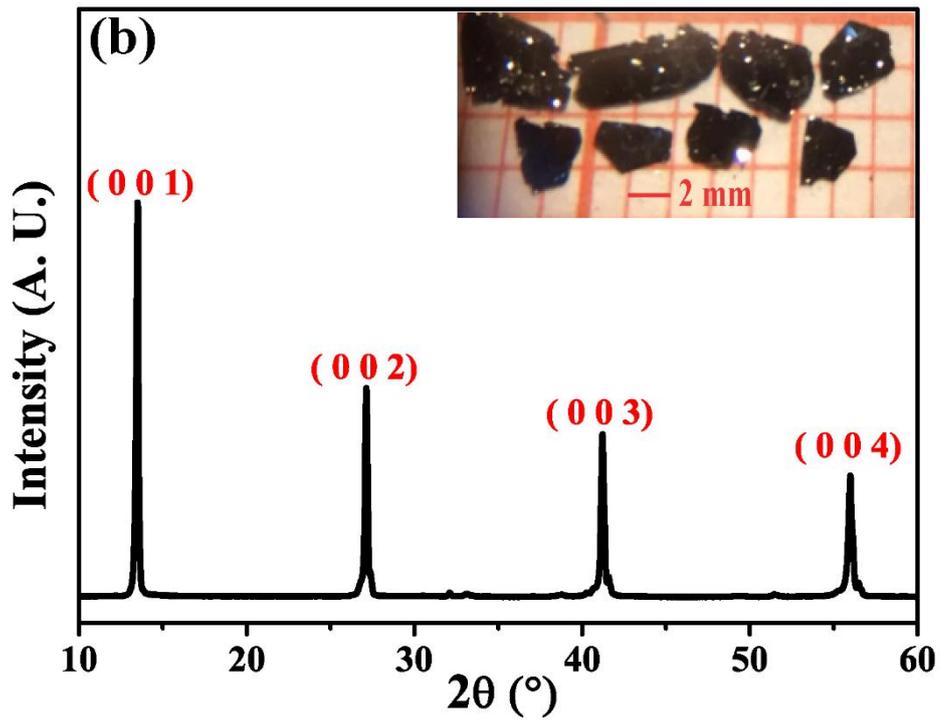

**Figure 2**. Powder XRD pattern with Rietveld refinement for 1T-NbSeTe (a), single crystal XRD pattern along (00l) (b).

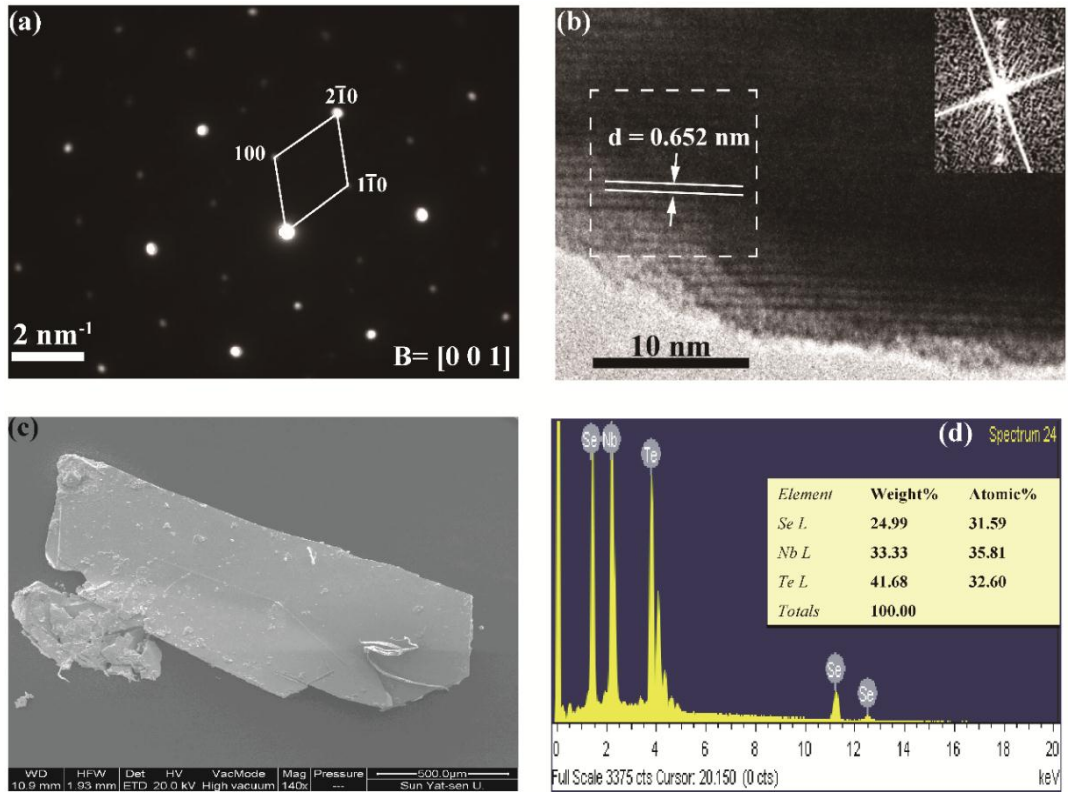

**Figure 3**. (a) Selected area electron diffraction pattern of NbSeTe polycrystalline powder taken on [001] orientation. (b) HRTEM image of NbSeTe polycrystalline powder. Inset is the corresponding FFT pattern. (c) SEM image of NbSeTe single crystal in the magnification of 140 times. (d) EDS spectrum and element ratio of NbSeTe.

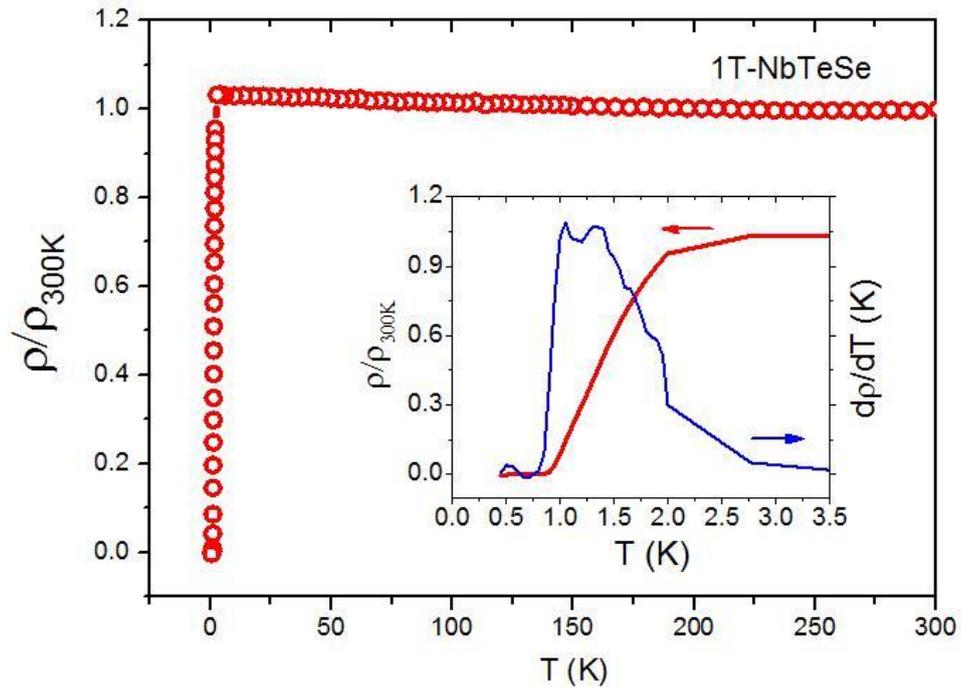

**Figure 4**. The temperature dependence electrical resistivity for polycrystalline NbSeTe. Inset shows the temperature dependence electrical resistivity at low temperature (red line) and metallic temperature dependence (d$\rho$/d$T$) in the temperature region of 0.5 - 3.5 K (blue line).

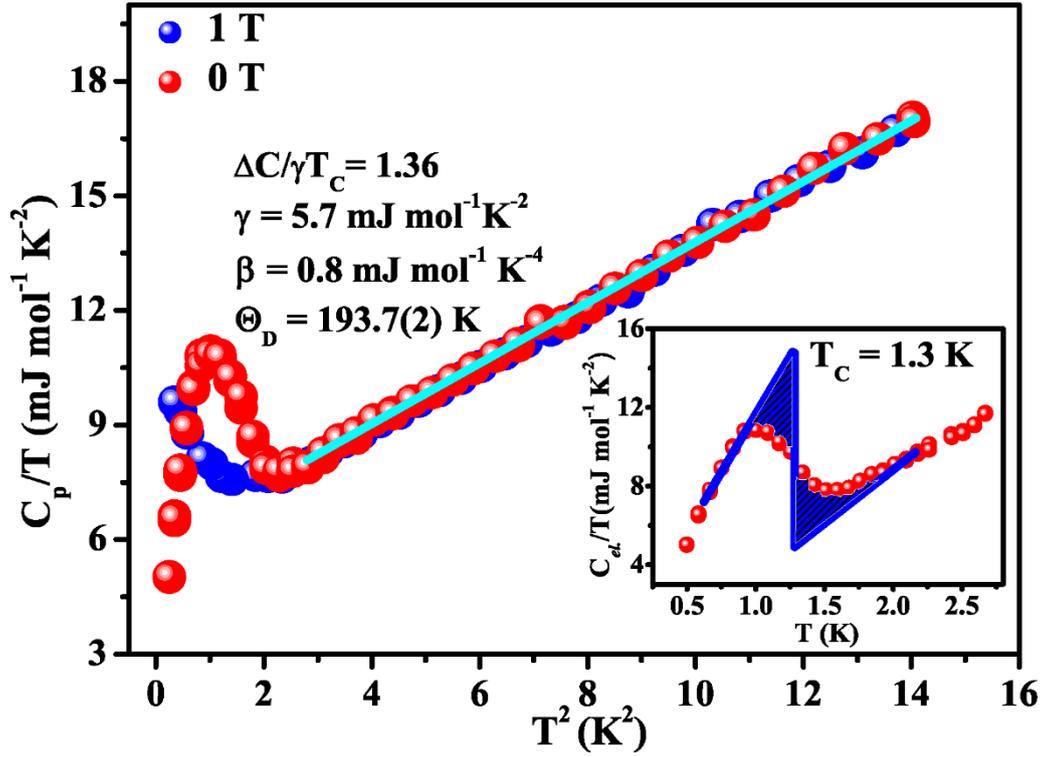

**Figure 5.** Heat Capacity characterization of 1T-NbSeTe. Debye temperature of 1T-NbSeTe obtained from fits to data in applied field of 1 T. Inset show heat capacities through the superconducting transitions without applied magnetic field for 1T-NbSeTe.

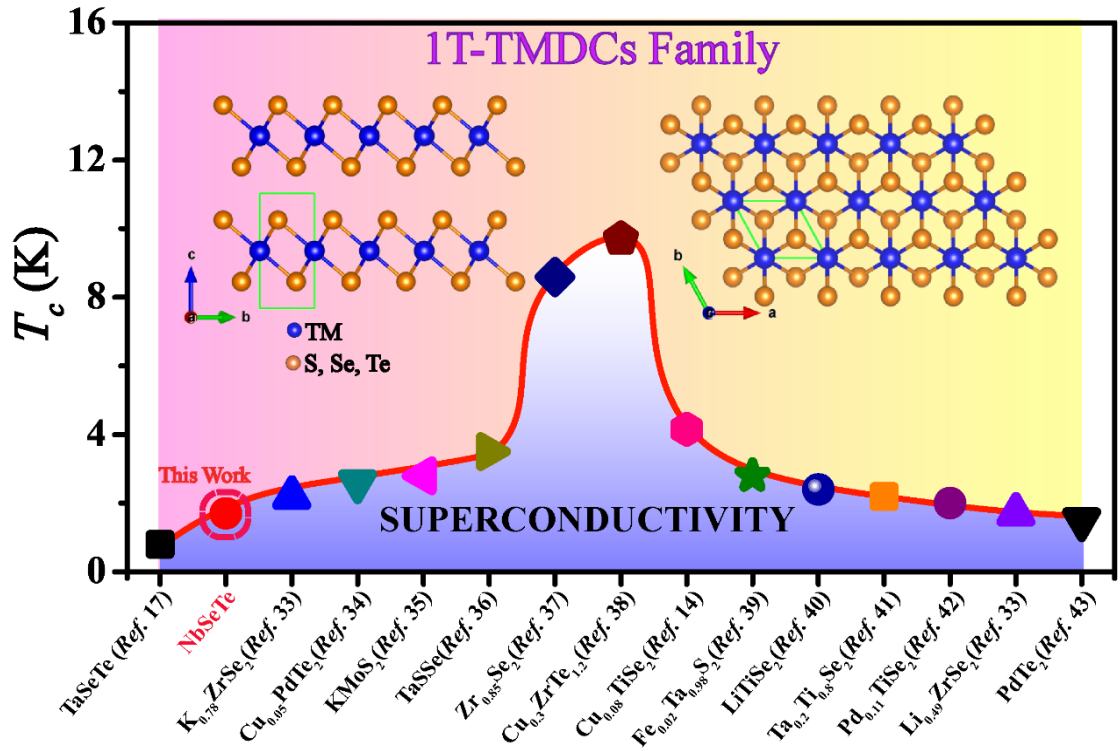

**Figure 6.** $T_c$ phase diagram of the 1T-TMDCs family.

## REFERENCES

1. Wilson, J. A.; Yoffe, A. D. The transition metal dichalcogenides discussion and interpretation of the observed optical, electrical and structural properties. *Adv. Phys.* **1969**, *18*, 193-335.

2. Friend, R. H.; Yoffe, A. D. Electronic properties of intercalation complexes of the transition metal dichalcogenides. *Adv. Phys.* **1987**, *36*, 1-94.

3. Lee, H. N. S.; McKinzie, H. Tannhauser, D. S.; Wold, A. The low-temperature transport properties of $NbSe_2$. *J. Appl. Phys.* **1969**, *40*, 602-604.

4. Yang, J. J.; Choi, Y. J.; Oh, Y. S.; Hogan, A.; Horibe, Y.; Kim, K.; Min, B. I.; Cheong, S. W. Charge-Orbital Density Wave and Superconductivity in the Strong Spin-Orbit Coupled $IrTe_2$: Pd. *Phys. Rev. Lett.* **2012**, *108*, 116402.

5. Rossnagel, K. On the origin of charge-density waves in select layered transition-metal dichalcogenides. *J. Phys.: Condens. Matter.* **2011**, *23*, 213001.

6. Gamble, F. R.; Osiecki, J. H.; Cais, M.; Pisharody, R. Intercalation complexes of lewis bases and layered sulfides: a large class of new superconductors. *Science* **1971**, *174*, 493-497.

7. Belopolski, I. Xu, S.; Ishida, Y.; Pan, X.; Yu, P.; Sanchez, D. S.; Zheng H.; Neupane, M.; Alidoust, N.; Chang, G.; Chang, T.; Wu, Y.; Bian, G.; Huang, S.; Lee, C.; Mou, D.; Huang, L.; Song, Y.; Wang, B.; Wang, G.; Yeh, Y.; Yao, N.; Rault, J. E.; Fevre, P. L.; Bertran, F.; Jeng, H.; Kondo, T.; Kaminski, A.; Lin, H.; Liu, Z.; Song, F.; Shin, S.; Hasan, M. Z. Fermi arc electronic structure and Chern numbers in the type-II Weyl semimetal candidate $Mo_xW_{1-x}Te_2$. *Phys. Rev. B* 2016, **94,** 085127.

8. Belopolski, I.; Sanchez, D. S.; Ishida, Y.; Pan, X.; Yu, P.; Xu, S.; Chang, G.; Chang, T.; Zheng, H.; Alidoust, N.; Bian, G.; Neupane, M.; Huang, S.; Lee, C.; Song, Y.; Bu, H.; Wang, G.; Li, S.; Eda, G.; Jeng, H.; Kondo, T.; Lin, H.; Liu, Z.; Song, F.; Shin, S.; Hasan, M. Z. Discovery of a new type of topological Weyl fermion semimetal state in $Mo_xW_{1-x}Te_2$. *Nat. Commun.* **2016**, *7*, 13643.

9. Zheng, H.; Xu, S.; Bian, G.; Guo, C.; Chang, G.; Sanchez, D. S.; Belopolski, I.; Lee, C.; Huang, S.; Zhang, X.; Sankar, R.; Alidoust, N.; Chang, T.; Wu, F.; Neupert, T.;